

\input phyzzx

%
\catcode`\@=11 
\def\papersize{\hsize=40pc \vsize=53pc \hoffset=0pc \voffset=1pc
   \advance\hoffset by\HOFFSET \advance\voffset by\VOFFSET
   \pagebottomfiller=0pc
   \skip\footins=\bigskipamount \normalspace }
\catcode`\@=12 
\papers
\def\to{\rightarrow}

\vsize=23.cm
\hsize=15.cm

\tolerance=500000
\overfullrule=0pt

\Pubnum={LPTENS-95/14 \cr
OUTP-95-11P\cr
{\tt hep-th@xxx/9503209} \cr
March 1995}

\date={}
\pubtype={}
\titlepage
\title{ON GRAVITATIONAL DRESSING OF \break
2D FIELD THEORIES IN CHIRAL
GAUGE} \author{Adel~Bilal}
\address{
CNRS - Laboratoire de Physique Th\'eorique de l'Ecole
Normale Sup\'erieure
\foot{{\rm unit\'e propre du CNRS, associ\'e \`a l'Ecole Normale
Sup\'erieure et l'Universit\'e Paris-Sud}}
 \nextline 24 rue Lhomond, 75231
Paris Cedex 05, France\break
{\tt bilal@physique.ens.fr}
}
\andauthor{
{Ian I. Kogan}
\foot{on leave of absence from
ITEP, Moscow, Russia}
}
\address{\it Theoretical Physics,
1 Keble Road, Oxford, OX1 3NP, U.K. \break
{\tt kogan@thphys.ox.ac.uk}
}

\abstract{After giving a pedagogical review of the chiral gauge approach to 2D
gravity, with particular emphasis on the derivation of the gravitational Ward
identities, we discuss  in some detail the interpretation of matter correlation
functions coupled to gravity in chiral gauge. We argue that in chiral gauge no
 {\it explicit} gravitational  dressing factor, analogue to
the Liouville exponential in conformal gauge, is necessary for
left-right symmetric matter operators. In particular, we examine
the gravitationally dressed four-point correlation function of products of left
and
right fermions.
We solve the
corresponding gravitational Ward
identity exactly: in the presence of gravity this four-point function exhibits
 a logarithmic short-distance
singularity, instead of the power-law singularity in the absence of gravity.
This
rather surprising effect is non-perturbative in the gravitational coupling and
is a
sign for logarithms in the gravitationally dressed operator product expansions.
We also discuss some perturbative evidence that the chiral Gross-Neveu
model may remain integrable when coupled to gravity.
 }

\endpage
\pagenumber=1

 \def\PL #1 #2 #3 {Phys.~Lett.~{\bf #1} (#2) #3}
 \def\NP #1 #2 #3 {Nucl.~Phys.~{\bf #1} (#2) #3}
 \def\PR #1 #2 #3 {Phys.~Rev.~{\bf #1} (#2) #3}
 \def\PRL #1 #2 #3 {Phys.~Rev.~Lett.~{\bf #1} (#2) #3}
 \def\CMP #1 #2 #3 {Comm.~Math.~Phys.~{\bf #1} (#2) #3}
 \def\IJMP #1 #2 #3 {Int.~J.~Mod.~Phys.~{\bf #1} (#2) #3}
 \def\JETP #1 #2 #3 {Sov.~Phys.~JETP.~{\bf #1} (#2) #3}
 \def\PRS #1 #2 #3 {Proc.~Roy.~Soc.~{\bf #1} (#2) #3}
 \def\IM #1 #2 #3 {Inv.~Math.~{\bf #1} (#2) #3}
 \def\JFA #1 #2 #3 {J.~Funkt.~Anal.~{\bf #1} (#2) #3}
 \def\LMP #1 #2 #3 {Lett.~Math.~Phys.~{\bf #1} (#2) #3}
 \def\IJMP #1 #2 #3 {Int.~J.~Mod.~Phys.~{\bf #1} (#2) #3}
 \def\FAA #1 #2 #3 {Funct.~Anal.~Appl.~{\bf #1} (#2) #3}
 \def\AP #1 #2 #3 {Ann.~Phys.~{\bf #1} (#2) #3}
 \def\MPL #1 #2 #3 {Mod.~Phys.~Lett.~{\bf #1} (#2) #3}

\def\d{\partial}
\def\dt{\partial_t}
\def\f{\phi}
\def\ix{\int {\rm d}^2 x \, }
\def\e{\epsilon}
\def\er{\epsilon_{\rm R}}
\def\b{\beta}
\def\m{\mu}
\def\n{\nu}
\def\r{\rho}
\def\l{\lambda}
\def\g{\gamma}
\def\G{\Gamma}
\def\xm{x^-}
\def\ym{y^-}
\def\zm{z^-}
\def\wm{w^-}

\def\xp{x^+}
\def\yp{y^+}
\def\zp{z^+}
\def\wp{w^+}

\def\tb{{\bar t}}
\def\D{\Delta}
\def\rmd{{\rm d}}

\def\la{\langle}
\def\ra{\rangle}

\def\P{\Psi}
\def\p{\psi}
\def\dd #1 #2{{\delta #1\over \delta #2}}

\def\rg{\ra\ra}
\def\h{h_{++}}
\def\rg{\sqrt{-g}}
\def\N{\nabla}
\def\hp{{\hat\p}}

{\bf \chapter{Introduction}}

What happens to a renormalizable two-dimensional field theory when it
is coupled two gravity? In general we don't know. Of course, during
the past years, tremendous progress has been made on a multitude of
particular models, either through discrete matrix model techniques or
in the continuum using the Liouville theory to describe gravity in
the conformal gauge. Often, as is the case for the Ising model in a
magnetic field, the coupling to gravity simplifies the theory
alowing for an exact solution otherwise not available. Not
withstanding these successes, the continuum methods are mainly
restricted to {\it conformal} field theories. Also, we only have a
very limited knowledge about correlation functions beyond the two- or
three-point functions. Recently, by studying general continuum {\it
non}-conformal field theories coupled to gravity in chiral gauge
\REF\KKP{I. Klebanov, I. Kogan and A. Polyakov, \PRL 71 1993 3243 .}
[\KKP] it was shown that the one-loop $\b$-function gets affected by
a universal gravitational factor.

We think that the chiral gauge approach deserves further exploration.
Here we report on some rather surprising results obtained in chiral
gauge concerning a four-point function. This note is organized as
follows: first, in section 2, we review the chiral gauge approach of Polyakov
et al.
\REF\HOU{A. Polyakov, in {\it Les Houches 1988: Fields, Strings  and
Critical Phenomena}, Elsevier Science Publishers 1989 .}
\REF\POL{A. Polyakov, \MPL A2 1987 893 .}
\REF\KPZ{V. Knizhnik, A. Polyakov and A. Zamolodchikov,
\MPL A3 1988 819 .}
[\HOU-\KPZ] to 2D gravity, hopefully putting it into a
pedagogical setting (see also ref.
\REF\CHAM{A.H. Chamseddine and M. Reuter, \NP B317 1989 757 .}
\CHAM). In
particular, we show how one obtains the gravitational action and the
gravitational Ward identities. Then, in section 3, we discuss in some detail
the
interpretation and relevance of {\it non-integrated} correlation
functions in chiral gauge. In particular, we point out why the {\it
non-integrated} chiral gauge two-point functions give the
gravitational scaling dimensions of the {\it integrated} conformal
gauge two-point functions. We further argue that, in chiral gauge, for
left-right symmetric matter operators $O_{\rm M}$ no {\it explicit}
gravitational
dressing factor, analogue of the Liouville exponential of conformal gauge, is
necessary, and that correlators like $\int {\rm d}^2x_1 \ldots  \int {\rm
d}^2x_n \la
O_{\rm M}(x_1)\ldots O_{\rm M}(x_n)\ra$ are well-defined quantities.
 Then, in section 4, we describe a sample computation of a
four-point function using the Ward identities. It is first obtained as
a perturbative solution of a partial differential equation. The
perturbation series diverges but can easily be resummed. The resulting
function, shown to be valid independent of perturbation theory, shows
strong non-perturbative  effects, namely logarithmic rather than
power-type short distance singularities, invalidating the
weak-coupling interpretation. Finally, in secttion 5, we discuss the relevance
of this
result to the gravitationally dressed operator product expansions, as
well as the implications for the gravitationnal dressing of the
integrable chiral Gross-Neveu model.

{\bf \chapter{ A review of 2D gravity in chiral gauge}}

\section{The matter action}

To start with, consider the action of a Majorana fermion $\chi$
coupled to gravity. Of course, one could consider a more general
matter action as well, but let's be specific.
$$S_{\rm M}={1\over \sqrt{2}}\ix (\det e) \bar\chi \g^a
e_{a}^{\phantom{a}\m}\d_\m\chi\ .
\eqn\di$$
Here $e_{a\m}$ is the zweibein and $ e_{a}^{\phantom{a}\m}$ its
inverse. Our conventions are fairly standard.\foot{
Our conventions are: $x^\pm={1\over \sqrt{2}}(x^0\pm x^1),\ \d_\pm=
{1\over \sqrt{2}}(\d_0\pm\d_1)$, $x_ay^a=x_+y^++x_-y^-$. The
Minkowski metric is $\eta_{00}=-1,\, \eta_{11}=1$ $ \Rightarrow\,
\eta_{+-}=\eta^{+-}=-1$, and $g_{\m\n}=e_{a\m}e^a_{\phantom{a}\n}$,
and $e_a^{\phantom{a}\m}e^a_{\phantom{a}\n}=\delta^\m_\n$ where, as
usual, Lorentz indices ($a,b,\ldots$) are raised and lowered with
$\eta_{ab}$ while $g_{\m\n}$ is used for curved space indices
($\m,\n,\ldots$). One defines $\bar\chi =\chi^+\g^0=\chi\g^0$ and
$\g^0\g^1=\g_5$ so that $\bar\chi\g^0=-\chi$ and
$\bar\chi\g^1=\chi\g_5$. Furthermore $(\g^0)^+=-\g^0, (\g^1)^+=\g^1,
\g_5^+=\g_5$ and $\chi_\pm={1\over 2} (1\pm\g_5)\chi$, so that
$\bar\chi\g^\pm={1\over
\sqrt{2}}\bar\chi(\g^0\pm\g^1)=-\sqrt{2}\chi_\mp$. Finally note that
for a $2\times 2$-matrix $M$ one has $\det
M=M_{00}M_{11}-M_{01}M_{10}=M_{++}M_{--}-M_{+-}M_{-+}$.
}
One finds
$$\eqalign{
S_{\rm M}&=\ix \left[ \chi_-(-e_{+-}\d_++e_{++}\d_-)\chi_- +
\chi_+(-e_{-+}\d_-+e_{--}\d_+)\chi_- \right]\cr
&=\ix \left[\p_-\left( \d_+-{e_{++}\over e_{+-}}\d_-\right) \p_- +
\p_+ \left( \d_--{e_{--}\over e_{-+}}\d_+\right)\p_+\right] \cr
}
\eqn\dii$$
where we have rescaled $\p_+=\sqrt{-e_{-+}}\chi_+$ and
$\p_-=\sqrt{-e_{+-}}\chi_-$ (the local Lorentz phase can be chosen
such that the square-roots are real). Note that although $\chi_+,
\chi_-$ are diffeomorphism scalars, $\p_+, \p_-$ bahave as
half-differentials.

If one now makes the conformal gauge choice $e_{++}=e_{--}=0$,
$e_{+-}=e_{-+}=-e^\f$ so that $g_{+-}=g_{-+}=-e^{2\f}$ and
$g_{++}=g_{--}=0$ one obtains the well-known free-fermion action in
conformal gauge:
$S_{\rm M}=\ix \left[ \p_- \d_+\p_- + \p_+ \d_-\p_+\right]$. Here, however, we
make a different gauge choice leading to the so-called chiral
gauge:
$$e_{+-} e_{-+}=1,\ e_{--}=0,\ {e_{++}\over e_{+-}}=\h
\eqn\diii$$
where the last equation is not a gauge choice but just the definition of $\h$.
Then the fermion action becomes
$$S_{\rm M}=\ix \left[ \p_- (\d_+-\h\d_-)\p_- + \p_+\d_-\p_+\right]\ .
\eqn\div$$
Gravity is represented by the field $\h$ and only the left fermion
$\p_-$ couples to gravity. It is straightforward to compute the
metric tensor in this gauge:
$$g_{+-}=g_{-+}=g^{+-}=g^{-+}=-1,\ g_{--}=g^{++}=0,\
g_{++}=-2\h,\ g^{--}=2\h
\eqn\dv$$
and $-g\equiv \det g_{\m\n}=1$. It is also easy to find that the only
non-vanishing Christoffel symbols are
$$\G^-_{+-}=\G^-_{-+}=\d_-\h,\ \G^+_{++}=-\d_-\h,\
\G^-_{++}=\d_+\h+2\h\d_-\h\ .
\eqn\dvi$$
The stress tensor is defined in general as
$$T_{\m\n}={1\over \rg} \dd S {g^{\m\n}} = {1\over \det e}\, e_{a\m}
\dd S {e_a^{\phantom{a}\n}}
\eqn\dvii$$
and one obtains for the matter part in chiral gauge
$$\eqalign{T^{\rm M}_{--}&=\p_-\d_-\p_-,\quad
T^{\rm M}_{+-}=T^{\rm M}_{-+}=\h\p_-\d_-\p_-,\cr
T^{\rm M}_{++}&=\p_+\d_+\p_++(\h)^2\p_-\d_-\p_- \ .\cr }
\eqn\dviii$$
It is straightforward to show that $T^{\rm M}_{\m\n}$ is classically
conserved: $\N^\m T^{\rm M}_{\m\n}=0$.
For $\n=+$ e.g., using \dvi\ this is equivalent to the vanishing
of
$$\d_+ T^{\rm M}_{-+}+\d_- T^{\rm M}_{++}
-(\d_+\h )T^{\rm M}_{--}-2(\d_-\h) T^{\rm M}_{+-}-2\h\d_- T^{\rm
M}_{-+}\ ,$$
which in turn is shown using the equations of motion
$$\d_+\p_-=\h\d_-\p_-+{1\over 2} (\d_-\h)\p_-\ ,\quad \d_-\p_+=0\ .
\eqn\dix$$
Note also that the stress tensor is traceless:
$(T^{\rm M})_\m^{\phantom{\m}\m}=0$.

\section{Diffeomorphisms}

Next, let us consider the effect of diffeomorphisms. Under a general
infinitesimal diffeomorphism $x^\pm\to {x'}^\pm=x^\pm+\e^\pm(x^+,x^-)$
one has
$$\eqalign{
\delta e_{a\m}(x)&\equiv e'_{a\m}(x) - e_{a\m}(x)
=\e^\l\d_\l e_{a\m} +e_{a\l}\d_\m \e^\l\cr
\delta g_{\m\n}(x)&\equiv g'_{\m\n}(x)-g_{\m\n}(x)=
\N_\m\e_\n+\N_\n\e_\m=g_{\n\r}\N_\m\e^\r+g_{\m\r}\N_\n\e^\r \cr
}
\eqn\dx$$
since $\N_\m g_{\n\r}=0$. Residual diffeomorphisms preserving chiral
gauge must obey $\delta g_{--}=\delta g_{+-}=0$ which implies (we
write $\er^\m$ for residual diffeomorphisms)
$$\N_\m\er^\m=\N_-\er^+=0 \quad \Leftrightarrow \quad
\d_\m\er^\m=\d_-\er^+=0 \ .
\eqn\dxi$$
It is a straightforward excercise to check, using
$\delta\p_\pm=\er^\l\d_\l\p_\pm \pm {1\over 2} (\d_+\er^+)\p_\pm$,
$\delta \h=\er^\l\d_\l\h+\d_+\er^-+2\h\d_+\er^+$, that the matter
action \div\ is invariant under residual diffeomorphisms. We will
also need to consider non-residual diffeomorphisms with general
$\e^-$, but $\e^+=0$. In this case one has to remember that $\h$ is
defined as $\h={e_{++}\over e_{+-}}$ while
$g_{++}=-2e_{++}e_{-+}=-2(e_{+-}e_{-+})\h$. Although $e_{+-}e_{-+}=1$
in chiral gauge, one has $\delta (e_{+-}e_{-+})=\d_-\e^-$ (but still
$\delta e_{--}=0$) so that $\delta g_{++}=-2 \d_-\e^-\h-2\delta\h$.
Thus %
$$\delta\h=\d_+\e^-+\e^-\d_-\h-\h\d_-\e^- \quad {\rm for}\ \e^+=0\ .
\eqn\dxii$$
Furthermore, for such diffeomorphisms
$$\eqalign{
\delta\p_+&=\e^-\d_-\p_+\cr
\delta\p_-&=\e^-\d_-\p_- +{1\over
2}(\d_-\e^-)\p_- \cr }\qquad {\rm for}\ \e^+=0
\eqn\dxiib$$
so that one can show that the matter action \div\ in
chiral gauge is still invariant under these non-residual
diffeomorphisms with $\e^+=0$.

\section{The gravitational action}

We want to compute the effective action $\G$ in chiral gauge, much
along the lines of Friedan's Les Houches lectures
\REF\FRI{D. Friedan, in {\it Les Houches 1982: Advances in Field
Theory and Statistical Mechanics}, Elsevier Science Publishers, 1984.}
for the conformal gauge [\FRI]. The matter part is rather trivial
since the original action is quadratic in the matter fields, but the
gravitational part will turn out to be much less trivial. Start with
the generating functional
$$ Z[\eta_+,\eta_-]={1\over \Omega_{\rm diff}}\int {\cal D}g{\cal D}\p
\exp\left( -S_0^{\rm gr}(g)-S_{\rm M}(g,\p) +\ix
\rg(\p_-\eta_-+\p_+\eta_+)\right)
\eqn\dxiii$$
where $S_0^{\rm gr}(g)=\m_0\ix\rg$ and $\Omega_{\rm diff}$ is the
volume of the diffeomorphism group. One has ${\cal D}g\equiv {\cal
D}g_{++}  {\cal D}g_{--} {\cal D}g_{+-}$ $= {\cal D}v^+ {\cal D}v^-
{\cal D}\h {\rm d}m \times J$ where $v^\pm$ are vector fields
parametrizing the diffeomorphisms, $m$ stands for the moduli, and the
Jacobian $J$ can be represented as usual by an integral over ghosts
[\POL]
$$\eqalign{J&\sim \int {\cal D}\xi_{++} {\cal D}\e_+ {\cal D}\zeta
{\cal D}\e_- e^{-S_{\rm gh}}\cr
S_{\rm gh}&=\ix \left[\xi_{++}\N_-\e_- +\zeta (\N_+\e_-+\N_-\e_+)
\right] \cr
}
\eqn\dxiv$$
so that
$$\eqalign{ Z[\eta_+,\eta_-]=&\int {\rm d}m\, {\cal D}\h\,
Z[\eta_+,\eta_-,\h,m]\cr
Z[\eta_+,\eta_-,\h,m]=&e^{-S_0^{\rm gr}}\left( \int{\cal D}{\rm gh}\,
e^{-S_{\rm gh}}\right)\left( \int{\cal D}\p\, e^{-S_{\rm M}}\right) \cr
&\times
\exp\left(-{1\over 4}\int\left(\eta_-
D_+^{-1}\eta_-+\eta_+\d_-^{-1}\eta_+\right)\right)  \cr }
\eqn\dxv$$
where $D_+=\d_+-\h\d_-$ and where we shifted as usual $\p_- \to\p_- +
{1\over 2} D_+^{-1}\eta_-$ and $\p_+\to\p_++{1\over
2}\d_-^{-1}\eta_+$. Let  $\Sigma_{\rm gh}(g)=-\log \int {\cal D}{\rm
gh}\, e^{-S_{\rm gh}}$ and $\Sigma_{\rm M}(g)=-\log \int {\cal D}\p\,
e^{-S_{\rm M}}$. Then the generating functional of connected
$\p$-correlation functions is\foot{
We do not explicitly write the dependence on the moduli any longer.}
$$\eqalign{
W[\eta_+,\eta_-,\h]&=\log Z[\eta_+,\eta_-,\h] \cr
&=-S_0^{\rm gr}
-\Sigma_{\rm gh}(\h) - \Sigma_{\rm M}(\h)
-{1\over 4}\int\left(\eta_-
D_+^{-1}\eta_-+\eta_+\d_-^{-1}\eta_+\right)\ .\cr }
\eqn\xvi$$
To obtain the effective action $\G$ one introduces the classical
fields $\hp_\pm=\dd W {\eta_\pm}$ so that
$$\eqalign{
\G[\hp_+,\hp_-,\h]&=\ix \left(
\eta_-\hp_-+\eta_+\hp_+\right) - W[\eta_+,\eta_-,\h]\cr
&=S_0^{\rm gr}
+\Sigma_{\rm gh}(\h) + \Sigma_{\rm M}(\h)
+\ix\left(\hp_-D_+\hp_-+\hp_+\d_-\hp_+\right)\cr
&\equiv \G_0[\h]+\G_{\rm excit}[\hp_+,\hp_-,\h]\cr }
\eqn\dxvii$$
which we seperated into a (non-trivial) ground-state contribution
$\G_0=S_0^{\rm gr}+\Sigma_{\rm gh} + \Sigma_{\rm M}$ and a trivial
excitation part $\G_{\rm excit}$ which is formally identical to the
original matter action $S_{\rm M}$. The effective action is to be
used to compute correlation functions as
$$\la \hp_\pm(x_1) \ldots \hp_\pm(x_n)\ra = \int {\rm d}m {\cal D}\h
{\cal D} \hp_+ {\cal D}\hp_-\  \hp_\pm(x_1) \ldots \hp_\pm(x_n)\,
e^{-\G[\hp_+,\hp_-,\h]}\ .
\eqn\dxviii$$

We have gone through the usual field theoretic formalism to show that
the gravitational part $\G_0$ of $\G$ gets contributions from the
ghost and matter sectors. Equivalently one could have argued {\it \`a
la} David-Distler-Kawai
\REF\DDK{F. David, \MPL A3 1988 1651 ; \nextline
J. Distler and H. Kawai, \NP B321 1989
509 .} [\DDK]  that in \dxv\ the measures ${\cal D}{\rm gh}
\equiv {\cal D}_g{\rm gh}$ and ${\cal D}\p\equiv {\cal D}_g \p$ are
complicated but factorize into $e^{-\Sigma_{\rm gh}}{\cal D}_0 {\rm
gh}$ and $e^{-\Sigma_{\rm M}} {\cal D}_0\p$ where ${\cal D}_0 {\rm
gh}$ and ${\cal D}_0\p$ now are trivial (flat) measures. Then one
would directly obtain \dxviii\ (except for the replacement $\hp_\pm
\to \p_\pm$).

Now, if the Faddeev-Popov procedure of factorizing the volume of the
diffeomorphism group is to make sense, then $\G$ must be invariant
under general diffeomorphisms, not only residual ones.\foot{
One may view this requirement as a condition on the counterterms.}
To check the invariance of $\G$ properly one should have imposed,
instead of \diii, conditions like $e_{+-}e_{-+}=\alpha$,
${e_{--}\over e_{-+}}=\beta$ and the definition ${e_{++}\over
e_{+-}}=\h$, and keep $\alpha$ and $\beta$ as {\it classical}
non-dynamical background fields. Then under a diffeomorphism \dx\ one
has e.g. $\delta\beta_{--}\big\vert_{\beta_{--}=0}=\d_-\e^+$. The
classical action $S_{\rm M}$ would then read
$$S_{\rm M}=\ix\left[\p_-(\d_+-\h\d_-)\p_-+\p_+(\d_--\beta_{--}\d_+)
\p_+\right]
\eqn\dxix$$
which is obviously invariant under diffeomorphisms, since it is just
the original  invariant matter action written in a particular form.
Actually, for our purpose of deriving the gravitational action $\G_0$
is is enough to consider diffeomorphisms with $\e^+=0$ and $\e^-$
arbitrary. In this case  there is no need to introduce the $\alpha$
and $\beta$ into $S_{\rm M}$ and one can directly work with $S_{\rm
M}$ as given by \div.
As already noted earlier, \div\ is invariant under diffeomorphisms
with $\e^+=0$. Since $S_{\rm M}$ is formally identical with $\G_{\rm
excit}$ the same is true for the latter. Thus imposing invariance of
$\G$ translates into imposing invariance of $\G_0[g_{\m\n}]$:
$$0=\delta \G_0=\ix \delta g^{\m\n} \dd {\G_0} { g^{\m\n}}
=\ix \rg 2\N^\m \e^\n T^{\rm grav}_{\m\n}
=-2\ix\rg \e^\n\N^\m T^{\rm grav}_{\m\n}
\eqn\dxx$$
where
$$T^{\rm grav}_{\m\n}={1\over \rg} \dd {\G_0} {g^{\m\n}} \ .
\eqn\dxxi$$
Equation \dxx\ implies the conservation of the gravitational
stress-energy tensor, in particular with $\e^+=0$:
$$\eqalign{
0&=\N^\m T^{\rm grav}_{\m -} = -\N_+ T^{\rm grav}_{--} + 2\h
\N_- T^{\rm grav}_{--} - \N_- T^{\rm grav}_{+-}\cr
&=\N_+ T^{\rm grav}_{--}+\h \N_- T^{\rm grav}_{--}
+{1\over 2} \N_-(T^{\rm grav})^\m_{\phantom{\m}\m}\cr }
\eqn\dxxii$$
or
$$\left[ \d_+-\h\d_--2(\d_-\h)\right] T^{\rm grav}_{--}
={1\over 2} \N_-(T^{\rm grav})^\m_{\phantom{\m}\m} \ .
\eqn\dxxiii$$
Now one uses the fact that $(T^{\rm grav})^\m_{\phantom{\m}\m}$ is a
gravitational scalar of dimension two, and hence must be proportional
to the curvature scalar:
$$ (T^{\rm grav})^\m_{\phantom{\m}\m} = {\l^g\over 24} R\ .
\eqn\dxxiv$$
Combining eqs \dxxiii\ and \dxxiv\ gives
$$\left[ \d_+-\h\d_--2(\d_-\h)\right] T^{\rm grav}_{--}
={\l^g\over 48} \d_- R
\eqn\dxxv$$
which is the central equation for determining the gravitational
action $\G_0$. It is straightforward to compute $R$ in chiral gauge:
$$R=2\d_-^3\h \ .
\eqn\dxxvi$$
Furthermore, since $\rg=1$ one has $T^{\rm grav}_{--}={1\over 2}
\dd {\G_0} {\h}$ and \dxxv\ turns into a functional differential
equation for $\G_0$:
$$\left[ \d_+-\h\d_--2(\d_-\h)\right] \dd {\G_0[\h]} {\h}
={\l^g\over 12} \d_-^3 \h \ .
\eqn\dxxvii$$
Its solution is
$$\G_0[\h]={\l^g\over 24}\ix (\d_-^2\h){1\over
\d_-(\d_+-\h\d_-)}\d_-^2\h + \m\ix \ .
\eqn\dxxviii$$
Indeed, observing that $(\d_-\d_+-\d_-\h\d_-)^{-1}$ is a symmetric
pseudodifferential operator one immediately finds
$$T^{\rm grav}_{--}={1\over 2}  \dd {\G_0} {\h}
={\l^g\over 24} \d_-^2{1\over  \d_+-\h\d_-}\d_-\h -
{\l^g\over 48}\left( \d_- {1\over  \d_+-\h\d_-}\d_-\h\right)^2 \ .
\eqn\dxxix$$
Applying then $\d_+-\h\d_-$ to the r.h.s. of this equation yields
${\l^g\over 24}\d_-^3\h + 2\d_-\h T^{\rm grav}_{--}$, showing that
\dxxvii\ is satisfied. Since $\d_-(\d_+-\h\d_-)$ is $-{1\over 2}$
times the Laplacian on a scalar, one finds of course that $\G_0$ can
be written as
$$\G_0=-{\l^g\over 48}\ix \rg R{1\over \N^2}R+\m\ix\rg
\eqn\dxxx$$
which is the famous covariant form written by Polyakov [\POL], and
which, in conformal gauge, reproduces the Liouville action.

\section{Gravitational Ward identities}

Note that equation \dxxvii\ can be rewritten, using \dxii, as
$$\ix \delta \h \dd {\G_0} {\h} =-{\l^g\over 12} \ix \e^-\d_-^3\h \ .
\eqn\dxxxi$$
This means that $\G_0$, if considered as a functional of $\h$ only, is
not (completely) invariant under diffeomorphisms (even with $\e^+=0$) but
has the anomaly as given by the r.h.s. of this equation. Equation
\dxx\ however, expresses the diffeomorphism invariance of $\G_0$ as a
covariant functional of $g_{++}$ and $g_{+-}$ (and $g_{--}$),
where the chiral gauge is only imposed after the variation is
performed.

It is now straightforward to derive Ward identities for correlation
functions:
$$\la\p(x_1) \ldots \p(x_n)\ra=\int {\cal D}\h e^{-\G_0[\h]}
\int {\cal D}\p\, \p(x_1) \ldots \p(x_n) e^{-\G_{\rm excit}[\h,\p]} \ .
\eqn\dxxxii$$
Here $\p$ need not be the previous fermion fields, but could be  more
general matter fields. Changing variables of integration to
$\tilde\h=\h+\delta\h$ and $\tilde\p=\p+\delta\p$ such that
$\delta\h$ and $\delta\p$ correspond to the diffeomorphisms \dxii\ and
\dxiii, hence $\delta\G_{\rm excit}=0$, one
obtains, using \dxxxi, the following Ward identity
$$\sum_{i=1}^n \la\p(x_1)\ldots \delta\p(x_i)\ldots \p(x_n)\ra
+{\l^g\over 12}\int {\rm d}^2 z\, \e^-(z) \la \d_-^3\h(z) \p(x_1)
\ldots \p(x_n)\ra =0 \ .
\eqn\dxxxiii$$

Next, following ref. \HOU, we will turn this Ward identity
into a partial differential equation for
$\la\p(x_1)\ldots\p(x_n)\ra$. To do so one has to eliminate $\h$ from
the correlation function. Here, we will again concentrate on the
above example of fermions. The classical equation of motion
$$\d_+\p_-=\h\d_-\p_-+\D (\d_-\h)\p_-
\eqn\dxxxiv$$
 with $\D={1\over 2}$ carries
over to the quantum case but for two differences: first, the product
of fields at the same point needs to be regularized by some
normal-ordering prescription which may modify the weight $\D$.
Second, when inserted into the functional integral, the equations of
motion remain true up to contact terms (e.g. $\la \dd S {\phi(x)}
\phi(y)\ra = \delta(x-y)$). Covariance of the quantum equation of
motion \dxxxiv\ requires that the variation of $\p_-$ under
diffeomorphisms also gets modified: eq. \dxiib\ gets replaced by
$$\delta \p_-=\e^-\d_-\p_-+\D(\d_-\e^-)\p_-
\eqn\dxxxv$$
with the same $\D$ as in \dxxxiv. The Ward identity then becomes
$$\eqalign{
&\sum_{i=1}^n \left( \delta^{(2)}(z-x_i) {\d\over \d x_i^-}
-\D {\d\over \d z^-} \delta^{(2)}(z-x_i) \right)
\la \p_-(x_1)\ldots \p_-(x_n)\ra \cr
&+{\l^g\over 12}\, {\d^3\over \d (z^-)^3}
\la \h(z) \p_-(x_1)\ldots \p_-(x_n)\ra =0 \ .\cr
}
\eqn\dxxxvi$$
Using the identities
$$\eqalign{
{\d^3\over \d (z^-)^3} {(z^--x_i^-)^2\over z^+-x_i^+} &=
4\pi i\, \delta^{(2)}(z-x_i) \ ,\cr
{\d^3\over \d (z^-)^3} {2(z^--x_i^-)\over z^+-x_i^+} &=
4\pi i\, {\d\over \d z^-} \delta^{(2)}(z-x_i) \ ,\cr
}
\eqn\dxxxvii$$
eq. \dxxxvi\ can be integrated as
$$\eqalign{
&\sum_{i=1}^n \left( {(z^--x_i^-)^2\over z^+-x_i^+}{\d\over \d x_i^-}
-2\D {(z^--x_i^-)\over z^+-x_i^+}\right)
\la \p_-(x_1)\ldots \p_-(x_n)\ra \cr
&=-{i\pi\l^g\over 3}
\la \h(z) \p_-(x_1)\ldots \p_-(x_n)\ra  \ .\cr
}
\eqn\dxxxviii$$
Now one uses the quantum equations of motion \dxxxiv\ as:
$$\eqalign{
&{\d\over \d z^+}\la \p_-(z)\p_-(x_2)\ldots \p_-(x_n)\ra \cr
&= \la \left(\h(z)\d_-\p_-(z)+\D\d_-\h(z)\p_-(z)\right) \p_-(x_2)\ldots
\p_-(x_n)\ra \cr
&={\d\over \d z^-}\la \h(z)\p_-(z)\p_-(x_2)\ldots \p_-(x_n)\ra
+(\D-1)\la (\d_-\h(z))\p_-(z)\p_-(x_2)\ldots \p_-(x_n)\ra \cr  }
\eqn\dxxxix$$
(up to contact terms). Using \dxxxviii\ and its derivative w.r.t.
$z^-$ one finally arrives at
$$\eqalign{
\left\{ \g{\d\over \d \zp}+\sum_{i=2}^n \left[ {(\zm-x_i^-)^2\over
\zp-x_i^+} {\d\over \d\zm}{\d\over \d x_i^-} +2\D {\zm-x_i^-\over
\zp- x_i^+} \left({\d\over \d x_i^-}-{\d\over \d\zm}\right)
-{2\D^2\over \zp-x_i^+}
\right]\right\}&\cr
\la \p_-(x_1)\ldots \p_-(x_n)\ra= 0& \cr}
\eqn\dxxxx$$
where we set
$$\g={i\pi\l^g\over 3}\ .
\eqn\dxxxxi$$
We quote without proof [\POL] that $\g$ is related to the
total central charge $c$ of the matter coupled to
gravity (e.g. $c={1\over 2}$ for a Majorana fermion) by the relation
$$\g={1\over 12}\left( c-13-\sqrt{(c-1)(c-25)}\right) \ .
\eqn\dxxxxia$$
Equation \dxxxx\ was first written in ref. \KKP.

As an example for the use of the Ward identities, consider the
two-point function of $\p_-$. Perturbation theory\foot{
One can do a simple Feynman diagram expansion of the two-point
function using the vertices and propagators derived from $\G$. This
is a perturbation series in ${1\over \l^g}\sim {1\over \g}$.}
suggests the ansatz
$$\la\p_-(x)\p_-(y)\ra\sim
{1\over (x^--y^-)^{1+2\delta} (x^+-y^+)^{2\delta} }
={\left[(x^--y^-)(x^+-y^+)\right]^{-2\delta} \over x^--y^-} \ .
\eqn\dxxxxii$$
Gravity only contributes a left-right symmetric factor
$\left[(x^--y^-)(x^+-y^+)\right]^{-2\delta}$. Furthermore, by the
usual arguments, ${1\over 2}+\delta$ must coincide with the anomalous
dimension $\D$ of eq. \dxxxv. Inserting this ansatz into \dxxxx\ for
$n=2$ gives an algebraic equation for $\D$:
$$\D-\D_0={\D(\D-1)\over \g}
\eqn\dxxxxiii$$
where $\D_0={1\over 2}$. This is the well-known KPZ-equation [\KPZ]
expressing the anomalous dimension $\D$ in the presence of gravity in
terms of the dimension $\D_0$ without gravity.

{\bf \chapter{Interpretation and relevance of non-integrated
correlation functions}}

What do correlation functions like \dxxxxii\ mean? Since one has
integrated over the metrics, i.e. over $\h$, what is the distance
between $x$ and $y$? Clearly, these are non-trivial questions. Let us
compare with what one does in conformal gauge. In conformal gauge,
one usually computes integrated correlation functions at fixed area
$A$, like
$$\la \ix e^{\alpha\f(x)} O(x)\int {\rm d}^2y\, e^{\alpha\f(y)} O(y)
\ra \Big\vert_{{\rm fixed} A}
\eqn\ti$$
where $O$ is a left-right symmetric matter
 field of conformal dimensions
$(\D_0,\D_0)$ (e.g. the product of our fermion fields $\p_-$ and
$\p_+$ with $\D_0={1\over 2}$) and $\f$ the Liouville field.
The constant $\alpha$ is chosen such that the  conformal
dimension of $e^{\alpha\f}$ is $(1-\D_0,1-\D_0)$, so that the total
integrand has conformal  dimensions $(1,1)$, and the integral is
invariant under conformal transformations. The area of the surface
can be fixed by adjusting the zero-mode of the Liouville field $\f$.
Correlation functions like \ti\ are conformal scalars, i.e. are
invariant under the {\it residual} diffeomorphisms of conformal gauge
 and have a well-defined
meaning. It has been shown [\DDK] that \ti\ scales with the area as
$A^{2-2\D}$ where the gravitational scaling dimension $\D$ is given
by the KPZ formula \dxxxxiii. Hence this $\D$ coincides with the
$\D$-exponent characterizing  the {\it non}-integrated two-point
function in chiral gauge:
$$\la O(x) O(y)\ra \sim {1\over (x^--y^-)^{2\D}(x^+-y^+)^{2\D-2\D_0}}
\times {1\over (x^+-y^+)^{2\D_0} }
\eqn\tii$$
where e.g. in the case of the fermions the first factor comes from
$\p_-$ and the second from $\p_+$. One sees that although such
non-integrated correlation functions in chiral gauge are not
invariant, their singularity structure (exponent $\D$) nevertheless
has an invariant meaning. Let us try to understand why $\D$ as given
by \tii\ should coincide with the gravitational scaling dimension of
\ti. In chiral gauge, since $\rg=1$, the area of a surface is
completely independent of the metric $\h$. Whereas in conformal gauge
one could choose the range of the coordinates $x^+, x^-$ to be fixed,
in chiral gauge their range is relevant to the geometry (and is part
of the moduli of the surface). Integrating \tii\ in $x$ and $y$ over
the surface then gives $A^{2-2\D}$, possibly up to an $A$-independent
constant. Thus the $\D$ characterizing the power-law behaviour of the
non-integrated two-point function directly gives the gravitational
scaling dimension  {\it without} further dressing by some field
$f(\h)$ that would be the chiral gauge analogue of the $e^{\alpha
\f}$-dressing. To understand why no such extra dressing is required
in chiral gauge, let's go back to the example of the fermion fields,
i.e. $O=\p_+\p_-$. Under an $\e^-$-diffeomorphism we had
$$\eqalign{
\delta \p_-&=\e^-\d_-\p_-+\D(\d_-\e^-)\p_-\cr
\delta \p_+&=\e^-\d_-\p_+\cr }
\eqn\tiii$$
and similarly one finds for an $\e^+$-diffeomorphism
$$\eqalign{
\delta \p_-&=\e^+\d_+\p_-+(\D-{1\over 2})(\d_+\e^+)\p_-\cr
\delta \p_+&=\e^+\d_+\p_++{1\over 2}(\d_+\e^+)\p_+\cr }
\eqn\tiv$$
which combines into
$$\eqalign{
\delta \p_-&=\e^\l\d_\l\p_-+\D(\d_\l\e^\l)\p_-
-{1\over 2}(\d_+\e^+)\p_-\cr
\delta \p_+&=\e^\l\d_\l\p_+ + {1\over 2}(\d_+\e^+)\p_+\ .\cr }
\eqn\tv$$
For residual diffeomorphisms (preserving chiral gauge) one has
$\d_\l\er^\l=0$ and it follows
$$\delta_{\rm R}\p_\pm=\er^\l\d_\l\p_\pm
\pm{1\over 2}(\d_+\e^+)\p_\pm
\eqn\tvi$$
and as a consequence
$$\delta_{\rm R}(\p_+\p_-)=\er^\l\d_\l(\p_+\p_-) =
\d_\l(\er^\l\p_+\p_-)
\eqn\tvii$$
which is a total derivative. Hence $\ix O(x)=\ix \p_+(x)\p_-(x)$ is
invariant under residual diffeomorphisms, which is the analogue of the
$(1,1)$ condition in conformal gauge. We see that no extra
gravitational factor $f(\h)$ is needed to achieve this, explaining
why the exponent $\D={1\over 2}+\delta$ of \dxxxxii\ directly gives
the gravitational scaling dimension.

This result is of course not
restricted to fermions. From \tiii-\tvii\  we see that if under a
general diffeomorphism
$$\delta O=\e^\l\d_\l O+\D(\d_\l\e^\l) O
\eqn\tviii$$
then under a residual diffeomorphism
$$\delta_{\rm R} O=\er^\l\d_\l O=\d_\l(\er^\l O)
\quad\Rightarrow\quad \delta_{\rm R}\ix O =0 \ .
\eqn\tix$$
Thus integrated $n$-point functions
$\int {\rm d}^2 x_1\ldots {\rm d}^2 x_n \la O(x_1)\ldots O(x_n) \ra$
are perfectly well-defined\foot{
Let us insist that well-defined means invariance (or covariance)
under residual diffeomorphisms. Since one works with a fixed gauge
this is as much as one can demand. Of course, one should be able to
compare to another, say conformal gauge, but this is beyond the scope
of this note.}
objects. Of course, it is just as meaningful to consider also the
non-integrated $n$-point functions which are scalar densities w.r.t.
residual diffeomorphisms.   The invariance of the integrated
$n$-point functions under residual diffeomorphisms should be
expressible as BRST-invariance.\foot{
Recall that in conformal gauge, BRST-invariance is essentially the
statement that the integrand is a $(1,1)$ field w.r.t. conformal
transformations, i.e. residual diffeomorphisms.}
To our knowledge, the BRST operator for the chiral gauge has not yet
been constructed, nor its cohomoly been investigated, but it is clear
that correlators like $\la \int {\rm d}^2 x_1 \p_+(x_1)\p_-(x_1)
\ldots \int {\rm d}^2 x_n \p_+(x_n)\p_-(x_n)\ra$ should turn out to
be BRST-invariant.

{\bf \chapter{The fermion four-point function}}

As a non-trivial example of how to use the Ward-identity \dxxxx\ we
now compute the fermion four-point function. As discussed in the next
section, this is also of some relevance to the gravitational dressing
of the chiral Gross-Neveu model. Therefore we also  add some colour
indices $i,j =1, \ldots, N$, and take as the matter action $N$ copies
of \div. We want to compute, for $i\ne j$
$$\eqalign{
G_4(w,x,y,z)&= \la\p_-^i(w)\p_+^i(w) \p_-^i(x)\p_+^i(x)
\p_-^j(y)\p_+^j(y) \p_-^j(z)\p_+^j(z)\ra \cr
&={1\over (w^+-x^+)(y^+-z^+) }
\la \p_-^i(w) \p_-^i(x) \p_-^j(y) \p_-^j(z)\ra \ .\cr }
\eqn\qi$$
In analogy with the two-point function we use the following ansatz
$$
\la \p_-^i(w) \p_-^i(x) \p_-^j(y) \p_-^j(z)\ra =
{f(t^-,t^+)\over (\wm-\xm)^{2\D} (\wp-\xp)^{2\D-1}
(\ym-\zm)^{2\D}(\yp-\zp)^{2\D-1} }
\eqn\qii$$
where the
 anharmonic ratio $t\equiv t^-$ is given as
usual by
$$t\equiv t^-={(\wm-\ym)(\xm-\zm)\over (\wm-\zm)(\xm-\ym)}
\eqn\qiii$$
and similarly for $\tb\equiv t^+$. Inserting the ansatz \qii\ into
the Ward identity \dxxxx\ with $n=4$ leads after some algebra to\foot{
Performing the change of variables carefully also leads to contact terms
involving
$\delta^{(2)}(x-y)$ or $\delta^{(2)}(w-z)$. We will assume that $x\ne y$ and
$w\ne z$,
so that we can drop these contact terms. In particular, below  we will consider
$t\to 1$, i.e. $x\to w$ or $y\to z$ which is perfectly compatible with this
assumption.
}
$$\left[ \g\tb\d_{\tb} +{1-t\over 1-\tb}(\tb-t)\dt t\dt
+(1-4\D) t\dt +2\D^2 {t+1\over t-1}\right] f(t,\tb)=0\ ,
\eqn\qiv$$
Note that $\D=1$ reproduces the equation derived in ref. \KKP\ for
the four-current correlation function.
The ansatz
\qii\ is justified by the fact that we obtain an equation for
$f(t,\tb)$ involving only $t, \tb, \dt, \d_{\tb}$ and not $x, y, z$
or $w$ explicitly.

 The partial differential equation \qiv\ has many
solutions. To pick out the physical ones we have to compare with
perturbation theory in ${1\over \g}$. Looking at the effective action
$\G$ (cf. \dxvii, \dxxviii) it is clear upon rescaling
$\tilde\h=\sqrt{\g}\h$ that each interaction involving a $\h$ is
accompanied by at least one factor of ${1\over \sqrt{\g}}$. Doing a
standard perturbative Feynman diagram expansion, we obtain
$$f=1-{1\over 2\g} {t+1\over
t-1}\log t\tb +{\cal O}(1/\g^2) =1-{2\D^2\over \g} {t+1\over t-1}\log
t\tb +{\cal O}(1/\g^2)
\eqn\qv$$
since $\D={1\over 2}+{\cal O}(1/\g)$.

In principle, the partial differential equation \qiv\ could be solved
order by order in a perturbation series in ${1\over \g}$. In
practice, this leads to very complicated poly-log integrals already
at low orders.
 The main difficulty  is
 the factor $1/ (1-\tb)$. If one considers the vicinity of $t=1$,
this difficulty disappears, and \qiv\ becomes
$$\left[ \g\tb\d_{\tb} +(t-1)\dt t\dt
+(1-4\D) \dt + {4\D^2\over t-1}\right] f_1(t,\tb)=0
\eqn\qvi$$
where the subscript $1$ on $f$ is to remind us that $f_1\sim f$
only in the vicinity of $t=1$. This equation \qvi\ can be solved
exactly. Writing the solution as a perturbation series in ${1\over \g}$
(matching to \qv) gives
$$f_1(t,\tb)\equiv f_1(g)=\sum_{n=0}^\infty {[(2\D)_n]^2\over n!}\,
g^n\quad, \quad g=-{\log \tb\over \g (t-1)}
\eqn\qvii$$
where $a_n\equiv a(a+1)(a+2)\ldots (a+n-1)$. At each order in ${1\over \g}$
one can of course replace $\log \tb$ by $\log t\tb$ in $g$, as
suggested by \qv.\foot{
 Then $f_1$ is no longer an exact solution of
\qvi\ but an exact solution of another equation, differing from \qvi\
only by higher order terms in $(t-1)$.}
The series \qvii\ has zero radius of convergence, but its Borel
transform can be recognized as the hypergeometric function
$$B[f_1](u)=\sum_{n=0}^\infty {[(2\D)_n]^2\over n! n!}\, u^n
=F(2\D, 2\D,1;u)\ .
\eqn\qviii$$
The inverse Borel transform
$$ f_1(g)=\int_0^\infty {\rm d}v\, e^{-v} B[f_1](uv)
\eqn\qix$$
gives the resummed function $f_1(g)$ in
terms of a Whittaker function. Alternatively, one can directly
observe that \qvii\ coincides up to an overall factor with the
asymptotic expansion of a Whittaker function. Hence
\REF\GR{I.S. Gradshteyn and I.M. Rhyzhik, {\it Table of integrals,
series and products}, Academic Press, 1980.} [\GR]
$$f_1(g)=\left(-{1\over g}\right)^{2\D}\P(2\D,1;-{1\over g})
=\left(-{1\over g}\right)^{2\D-1/2}e^{-1/2g}\,
 W_{1/2-2\D,0}(-{1\over g})\ .
\eqn\qx$$
Here $W$ is the Whittaker function and $\P$ is a solution to the
degenerate hypergeometric equation [\GR]. Indeed, although the
equation \qvi\ has many solutions, if one uses an ansatz with $f_1$
only depending on $t$ and $\tb$ through $g$, then equation \qvi\
becomes
$$\left\{ g^2{\rmd^2\over \rmd g^2} +[(1+4\D)g-1] {\rmd\over \rmd g}
+4\D^2\right\} f_1(g) =0\ .
\eqn\qxi$$
Setting $f_1(g)=(-1/g)^{2\D} u(-1/g)$ one sees that $u(x)$ satisfies
the degenerate hypergeometric equation
$$xu''(x)+(b-x)u'(x)-a u(x)=0
\eqn\qxii$$
with $b=1$ and $a=2\D$. Perturbation theory has told us which of the
two independent solutions to choose, namely $u(x)=\P(2\D,1;x)$. Let
us insist that we just showed that \qx\ is a solution to the
differential equation \qvi, independent of perturbation
theory in ${1\over \g}$.

Having the perfectly non-perturbative expression \qx\ for $f_1(g)$,
we can now investigate its behaviour for large $g$, which is just the
series expansion of $\P(2\D,1;x)$ for small $x$:
$$f_1(g)=\left(-{1\over g}\right)^{2\D}\sum_{k=0}^\infty
{\G(2\D+k)\over [k! \G(2\D)]^2}
\left[ 2\p(k+1)-\p(2\D+k)-\log\left(-{1\over g}\right) \right]
\left(-{1\over g}\right)^k
\eqn\qxiii$$
where $\p(x)=\G'(x)/\G(x)$. Recall that $g=-\log\tb/[\g(t-1)]$, hence
large $g$ means $t\to 1$ (for fixed $\tb$), so this is the limit
where $f\sim f_1$. Although \qxiii\ is the exact asymptotic for
$f_1$, it gives only the leading order for $f$ (i.e. we can only trust the
$k=0$ term):
$$f(t,\tb)\sim \left( {\g(t-1)\over \log\tb}\right)^{2\D}
{1\over \G(2\D)} \left[ \p(1)-\log\left( {\g(t-1)\over \log\tb}\right)
+{\cal O}\left( (t-1), (t-1)\log (t-1)\right) \right] \ .
\eqn\qxiv$$

What does this mean for the fermion four-point function \qi?
Since
$$t-1={(\wm-\xm)(\ym-\zm)\over (\wm-\zm)(\xm-\ym)}\ ,
\eqn\qxv$$
one has
$t\to 1$ if either $\wm\to\xm$ or $\ym\to\zm$, i.e. when two fermion
operators of the same colour approach each other. Inserting \qxiv\
into \qi\ and \qii\ then gives
$$\eqalign{G_4(w,x,y,z)\sim&
{\g^{2\D}\over \G(2\D)}\left[
(\wm-\zm)(\xm-\ym)\right]^{-2\D}\left[\log\tb\right]^{-2\D}
\left[(\wp-\xp)(\yp-\zp)\right]^{-2\D}\cr
&\times \left[ \p(1)+\log\left({\log\tb\over \g}\right)
-\log {(\wm-\xm)(\ym-\zm)\over (\wm-\zm)(\xm-\ym)} \right] \ . \cr}
\eqn\qxvi$$
It is important to realize that we work in Minkowski space so that we
can take $t\to 1$, keeping $\tb\ne 1$ fixed. Rather surprisingly, the
four-point function \qxvi\ no longer contains the perturbative
singularity $\sim (\wm-\xm)^{-2\D} (\ym-\zm)^{-2\D}$, but resumming
the series has transformed it into a logarithmic singularity, plus a
non-singular part!\foot{
Naively it looks as if \qxvi\ now contains a new singularity $\sim
(\wm-\zm)^{-2\D} (\xm-\ym)^{-2\D}$ as $\wm\to\zm$ or $\xm\to\ym$.
However, this means $t-1\to\infty$ which is clearly outside the
domain of validity of eq. \qxvi.
Let us also insist that the contact terms $\sim\delta^{(2)}(x-y),\
\delta^{(2)}(x-y)$ we dropped above precisely correspond to $t-1\to\infty$ and
are
completely irrelevant to the behaviour as $t\to 1$.
}

Mathematically, the origin of the logarithm can be traced to the
degenerate hypergeometric equation \qxii\ satisfied by $u(x)=x^{-2\D}
f_1(-1/x)$. For generic parameter $b$ it has two independent
solutions [\GR] $\Phi(a,b;x)$ and $x^{1-b}\Phi(a+1-b,2-b;x)$.
Obviously, for $b\to 1$ the second solution generates $\log x\
\Phi(a,b;x)$, among others. This is a well-known phenomenon in the
theory of ordinary linear differential equations.

Physically however, it was quite unexpected that turning on gravity
(${1\over \g} \ne 0$), even infinitesimally weakly, completely changes the
singularity structure: this is a truely non-perturbative
phenomenon, due to the divergence of the perturbative series in
${1\over \g}$.

{\bf \chapter{Conclusions and Outlook}}

What do we learn from all these computations of the four-point
function? One obvious lesson is - unlike the situation of the
two-point function - that we cannot trust the weak-coupling
gravitational perturbation theory in ${1\over \g}\sim {1\over \l^g}$.
What is the meaning of the (non-perturbative) logarithmic singularity
of the four-point function? In section 3, we have argued that
integrating correlation functions like
\qxvi,  computed in chiral gauge, leads to well-defined objects,
invariant under residual diffeomorphisms. Obviously, we cannot
integrate our result \qxvi\ since it is valid only in the vicinity
$t\sim1$. However, since integrating is a ``trivial procedure" it
certainly makes sense to study the properties of the non-integrated
correlation functions as well. The main question that arises is
whether the correlator \qxvi\ tells us something about the
gravitationally dressed operator product expansion. Equation \qxvi\
should express the OPE of $\f(w)=\p_-(w)\p_+(w)$ with
$\f(x)=\p_-(x)\p_+(x)$ as $\wm\to\xm$. It looks like
$$\f(w)\f(x)\sim \tilde O(x)+\log(\wm-\xm) O(x) \ .
\eqn\ci$$
This is a particular example of a more general OPE with logarithmic
short-distance behaviour:
$$\f(w)\f(x)\sim\sum_n (w-x)^{\D_n-2\D_\f}\left[\tilde O_n+\ldots +
\log(w-x) O_n +\ldots \right]\ .
\eqn\cii$$

In conformal gauge, such logarithms have been noticed before in ref.
\REF\SAL{L. Rozansky and H. Saleur, \NP B376 1992 441 .}
\SAL\ where the WZW model based on the supergroup $Gl(1,1)$ was
discussed. Later a more systematic discussion was given in ref.
\REF\GUA{V. Gurarie, \NP B410 1993 535 .}
\GUA, where the appearance of logarithms in conformal blocks in the
$c=-2$ and other non-unitary models was studied. There it has been
argued that the emergence of logarithms in correlation  functions is
due to new so-called logarithmic operators, whose OPEs display
logarithmic short-distance singularities. These new logarithmic
operators have conformal dimensions degenerate with those of the
usual primary operators, and it is this degeneracy that is at the
origin of the logarithms (cf. our discussion of the degenerate
hypergeometric equation in the previous section). As a result one can
no longer completely diagonalize the Virasoro operator $L_0$, and the
new operators, together with the standard ones form the basis of the
Jordan-cell for $L_0$. In the case of two operators with degenerate
conformal dimensions $\D_n$ the operator product expansion precisely
takes the form \cii, while the OPE with the conformal stress-energy
tensor  is
$$T(z)\tilde O_n(0)\sim {\D_n\over z^2}\tilde O_n(0)+{1\over z^2}
O_n(0)+{1\over z}\d \tilde O_n(0)\ ,
\eqn\ciii$$
in particular
$$L_0\vert O_n\ra=\D_n\vert O_n\ra \ ,
\quad L_0 \vert \tilde
O_n\ra=\D_n\vert \tilde O_n\ra +\vert O_n\ra \ .
\eqn\civ$$
This makes it possible to have logarithmic terms in the correlation
functions without spoiling the conformal invariance.

Finally we would like to comment on the relevance of our present
results for the gravitational dressing of a two-dimensional
integrable but not conformally invariant field theory, namely the
chiral Gross-Neveu model
\REF\GN{D. Gross and A. Neveu, \PR D10 1974 3235 .}
[\GN]. As is well-known, its action is given by the massless
free-fermion action \di, where the fermions are $N$ component fields,
and an interaction term between two left and two right fermions $\sim
\ix \p_-^j\p_-^i\p_+^i\p_+^j$. In chiral gauge, only the left
fermions $\p_-$ interact with gravity, cf. eq. \div. Without gravity,
it is known that this model is completely integrable
\REF\LOW{N. Andrei and J.H. Lowenstein, \PRL 43 1979 1698 .}
[\LOW] and exhibits dynamical mass generation [\GN]. Does the
integrability remain once the model is coupled to gravity? A
necessary condition of integrability is that the $S$-matrix for the
scattering of the physical particles (here the massive fermions) is
factorizable and elastic. This actually is a consequence of the
factorizability and elasticity of the $S$-matrix for the
pseudoparticles (here the original massless fermions).

As the simplest
check, we have investigated whether the two-pseudoparticle $S$-matrix
remains elastic in the presence of gravity. Here again we face the
issue of the interpretation of the $S$-matrix elements in the
presence of gravity: the $S$-matrix, e.g. for the scattering of two
left fermions, is obtained from the four-point function \qii\ by
removing the external propagators and setting the external momenta
on-shell ($p_+=p'_+=q_+=q'_+=0$ where $p,p'$ and $q,q'$ are the
initial and final momenta). According to our discussion of section 3,
this does not seem to lead to a well-defined quantity. However,
bearing in mind the ``experimental" situation for measuring
$S$-matrix elements, even in the presence of gravity, we expect that
the $S$-matrix should be well-defined at least within a gravitational
weak-coupling expansion, provided the latter makes sense. Now for the
chiral Gross-Neveu model, there is no scattering of two right
pseudoparticles, while the left-right scattering is always elastic,
as can be seen simply by combining momentum conservation and the
on-shell condition. It remains to consider the scattering of two left
fermions which interact due to their coupling to gravity, cf. eq. \qii.
To first order in ${1\over \g}$, this $S$-matrix element vanishes.
Indeed, from \qv\ one finds upon Fourier transforming that is is given
by ${1\over \g}(p+p')_-(q+q')_-{(p-p')_+\over (p-p')_-^3}$ which
vanishes on-shell ($p_+=p'_+=0$). We have verified that this remains
true at the next order in ${1\over \g}$, including ``two graviton
exchanges". If this remains true at all orders in ${1\over \g}$ and
even non-perturbatively, one would have complete elasticity of the
two-pseudoparticle $S$-matrix, and this would certainly be a
well-defined and gauge-invariant statement. One could then go on and
speculate that all $S$-matrix elements for the pseudoparticle
scattering remain elastic and factorizable in the presence of gravity
and that the same is true for the physical (massive) fermions, in
other words that the integrability of the Gross-Neveu model survives
coupling to gravity. However, there is still a long way to go.

\ack

We are grateful to D. Gross, I. Klebanov and A. Polyakov for sharing
their insights at the earlier stage of this work when both authors
were still at Princeton University. One of us (A.B.) wishes to
acknowledge the hospitality of the Theory Division at CERN where this
work was completed.

\refout
\end